\documentclass[twocolumn,preprintnumbers,amsmath,amssymb]{revtex4}

\usepackage{graphicx}
\usepackage{amssymb,amsmath}

\begin{document}

\title{
Entanglement entropy of  integer Quantum Hall states
}

\author{Iv\'an D. Rodr\'{\i}guez and
Germ\'an Sierra}

\affiliation{Instituto de F\'{\i}sica Te\'orica,
UAM-CSIC, Madrid, Spain}

\bigskip\bigskip\bigskip\bigskip

%
\font\numbers=cmss12
\font\upright=cmu10 scaled\magstep1
\def\stroke{\vrule height8pt width0.4pt depth-0.1pt}
\def\topfleck{\vrule height8pt width0.5pt depth-5.9pt}
\def\botfleck{\vrule height2pt width0.5pt depth0.1pt}
\def\Zmath{\vcenter{\hbox{\numbers\rlap{\rlap{Z}\kern
0.8pt\topfleck}\kern 2.2pt
                   \rlap Z\kern 6pt\botfleck\kern 1pt}}}
\def\Qmath{\vcenter{\hbox{\upright\rlap{\rlap{Q}\kern
                   3.8pt\stroke}\phantom{Q}}}}
\def\Nmath{\vcenter{\hbox{\upright\rlap{I}\kern 1.7pt N}}}
\def\Cmath{\vcenter{\hbox{\upright\rlap{\rlap{C}\kern
                   3.8pt\stroke}\phantom{C}}}}
\def\Rmath{\vcenter{\hbox{\upright\rlap{I}\kern 1.7pt R}}}
\def\Z{\ifmmode\Zmath\else$\Zmath$\fi}
\def\Q{\ifmmode\Qmath\else$\Qmath$\fi}
\def\N{\ifmmode\Nmath\else$\Nmath$\fi}
\def\C{\ifmmode\Cmath\else$\Cmath$\fi}
\def\R{\ifmmode\Rmath\else$\Rmath$\fi}
\def\H{{\cal H}}
\def\NN{{\cal N}}
\def\tv{{\tilde{v}}}
\def\vep{{\tilde{\epsilon}}}
\def\te{{\tilde{\vep}}}
\def\sh{{\rm sh}}
\def\cth{{\rm cth}}
\def\th{{\rm th}}
\def\bk{{\bf k}}
\def\br{{\bf r}}
\def\Erf{{\rm Erf}}
\def\Erfc{{\rm Erfc}}

\begin{abstract}
We compute the entanglement entropy, in real space, of the ground state
of the integer Quantum Hall states for three different domains
embedded in  the torus, the  disk and  the sphere. We establish the validity of the area law with a
vanishing value of the topological entanglement entropy.
The entropy per unit length of the perimeter
depends on the filling fraction,  but it is independent of the geometry.
\end{abstract}


\maketitle

\vskip 0.2cm

%
%
%
%
\def\oti{{\otimes}}
\def\lb{ \left[ }
\def\rb{ \right]  }
\def\tilde{\widetilde}
\def\bar{\overline}
\def\hat{\widehat}
\def\*{\star}
\def\[{\left[}
\def\]{\right]}
\def\({\left(}      \def\BL{\Bigr(}
\def\){\right)}     \def\BR{\Bigr)}
    \def\BBL{\lb}
    \def\BBR{\rb}
%
%
\def\zb{{\bar{z} }}
\def\zbar{{\bar{z} }}
\def\frac#1#2{{#1 \over #2}}
\def\inv#1{{1 \over #1}}
\def\half{{1 \over 2}}
\def\d{\partial}
\def\der#1{{\partial \over \partial #1}}
\def\dd#1#2{{\partial #1 \over \partial #2}}
\def\vev#1{\langle #1 \rangle}
\def\ket#1{ | #1 \rangle}
\def\rvac{\hbox{$\vert 0\rangle$}}
\def\lvac{\hbox{$\langle 0 \vert $}}
\def\2pi{\hbox{$2\pi i$}}
\def\e#1{{\rm e}^{^{\textstyle #1}}}
\def\grad#1{\,\nabla\!_{{#1}}\,}
\def\dsl{\raise.15ex\hbox{/}\kern-.57em\partial}
\def\Dsl{\,\raise.15ex\hbox{/}\mkern-.13.5mu D}
%
%
\def\ga{\gamma}     \def\Ga{\Gamma}
\def\be{\beta}
\def\al{\alpha}
\def\ep{\epsilon}
\def\vep{\varepsilon}
\def\dep{d}
\def\arc{{\rm Arctan}}
\def\la{\lambda}    \def\La{\Lambda}
\def\de{\delta}     \def\De{\Delta}
 \def\hD{{\Delta}}
\def\om{\omega}     \def\Om{\Omega}
\def\sig{\sigma}    \def\Sig{\Sigma}
\def\vphi{\varphi}
\def\sign{{\rm sign}}
\def\he{\hat{e}}
\def\hf{\hat{f}}
\def\hg{\hat{g}}
\def\ha{\hat{a}}
\def\hb{\hat{b}}
%
%
\def\CA{{\cal A}}   \def\CB{{\cal B}}   \def\CC{{\cal C}}
\def\CD{{\cal D}}   \def\CE{{\cal E}}   \def\CF{{\cal F}}
\def\CG{{\cal G}}   \def\CH{{\cal H}}   \def\CI{{\cal J}}
\def\CJ{{\cal J}}   \def\CK{{\cal K}}   \def\CL{{\cal L}}
\def\CM{{\cal M}}   \def\CN{{\cal N}}   \def\CO{{\cal O}}
\def\CP{{\cal P}}   \def\CQ{{\cal Q}}   \def\CR{{\cal R}}
\def\CS{{\cal S}}   \def\CT{{\cal T}}   \def\CU{{\cal U}}
\def\CV{{\cal V}}   \def\CW{{\cal W}}   \def\CX{{\cal X}}
\def\CY{{\cal Y}}   \def\CZ{{\cal Z}}

\def\Hp{{\mathbb{H}^2_+}}
\def\Hm{{\mathbb{H}^2_-}}

\def\rvac{\hbox{$\vert 0\rangle$}}
\def\lvac{\hbox{$\langle 0 \vert $}}
\def\comm#1#2{ \BBL\ #1\ ,\ #2 \BBR }
\def\2pi{\hbox{$2\pi i$}}
\def\e#1{{\rm e}^{^{\textstyle #1}}}
\def\grad#1{\,\nabla\!_{{#1}}\,}
\def\dsl{\raise.15ex\hbox{/}\kern-.57em\partial}
\def\Dsl{\,\raise.15ex\hbox{/}\mkern-.13.5mu D}
%
%
%
\font\numbers=cmss12
\font\upright=cmu10 scaled\magstep1
\def\stroke{\vrule height8pt width0.4pt depth-0.1pt}
\def\topfleck{\vrule height8pt width0.5pt depth-5.9pt}
\def\botfleck{\vrule height2pt width0.5pt depth0.1pt}
\def\Zmath{\vcenter{\hbox{\numbers\rlap{\rlap{Z}\kern
0.8pt\topfleck}\kern 2.2pt
                   \rlap Z\kern 6pt\botfleck\kern 1pt}}}
\def\Qmath{\vcenter{\hbox{\upright\rlap{\rlap{Q}\kern
                   3.8pt\stroke}\phantom{Q}}}}
\def\Nmath{\vcenter{\hbox{\upright\rlap{I}\kern 1.7pt N}}}
\def\Cmath{\vcenter{\hbox{\upright\rlap{\rlap{C}\kern
                   3.8pt\stroke}\phantom{C}}}}
\def\Rmath{\vcenter{\hbox{\upright\rlap{I}\kern 1.7pt R}}}
\def\Z{\ifmmode\Zmath\else$\Zmath$\fi}
\def\Q{\ifmmode\Qmath\else$\Qmath$\fi}
\def\N{\ifmmode\Nmath\else$\Nmath$\fi}
\def\C{\ifmmode\Cmath\else$\Cmath$\fi}
\def\R{\ifmmode\Rmath\else$\Rmath$\fi}

\def\barray{\begin{eqnarray}}
\def\earray{\end{eqnarray}}
\def\beq{\begin{equation}}
\def\eeq{\end{equation}}

\def\no{\noindent}

\def\gpar{g_\parallel}
\def\gperp{g_\perp}

\def\Jb{\bar{J}}
\def\dx{\frac{d^2 x}{2\pi}}

\def\rap{\beta}
\def\s{\sigma}
\def\spec{\zeta}
\def\comb{\frac{\rap\theta}{2\pi} }
\def\Ga{\Gamma}

\def\L{{\cal L}}
\def\g{{\bf g}}
\def\K{{\cal K}}
\def\I{{\cal I}}
\def\M{{\cal M}}
\def\F{{\cal F}}

\def\gpar{g_\parallel}
\def\gperp{g_\perp}
\def\Jb{\bar{J}}
\def\dx{\frac{d^2 x}{2\pi}}
\def\imag{\Im {\it m}}
\def\real{\Re {\it e}}
\def\Jbar{{\bar{J}}}
\def\kh{{\hat{k}}}
\def\Im{{\rm Im}}
\def\Re{{\rm Re}}
\def\ti{{\tilde{\phi}}}
\def\tR{{\tilde{R}}}
\def\tS{{\tilde{S}}}
\def\tF{{\tilde{\cal F}}}
\def\ba{\bar{a}}
\def\bb{\bar{b}}
\def\be{\bar{\vep_0}}
\def\bD{\bar{\Delta_0}}

In recent years the notion of entanglement
has become a new tool for analyzing
the quantum states that arise in Condensed
Matter systems \cite{amico}. This notion has brought a
quantum information perspective to traditional
problems and techniques in the field, such as
quantum phase transitions, numerical simulation
methods, renormalization group, etc. A generic
measure of entanglement is given by the von Neumann
entropy $S_A$ of the reduced density matrix of a part $A$
of the total system. This quantity measures the amount
of quantum entanglement of the subsystem $A$
with its environment, usually denoted as $B$. For finite
systems, one has that $S_A = S_B$, so this quantity
reflects a property shared by $A$ and $B$.
If the quantum
state has a finite correlation length, then
heuristic arguments implies that the entropy is proportional to the area
of the common boundary between $A$ and $B$. This statement
is known as the area law, and it seems to be a universal
property satisfied by the quantum states appearing
in Nature  (see  \cite{cirac} for some
rigorous results). There are logarithmic  violations of this law
in critical one dimensional systems,
and some higher dimensional fermionic systems,
but the former ones can be understood using Conformal Field Theory \cite{cardy}.

Particularly interesting are the two dimensional
systems with topological order, where the entropy
law becomes $S_A = c L - \gamma + \CO(1/L)$,
where $L$ is the length of the boundary,
$c$ is a non universal constant
 and $\gamma$
is a quantity called topological entanglement entropy \cite{kitaev}.
The excitations of these  systems
are anyons and it turns out that $\gamma$
is  the logarithm of the total quantum dimension of these anyons.
The paradigmatic system with (abelian) anyons
is the Fractional Quantum Hall (FQH) state with filling
fraction $1/m$, for which $\gamma = \frac{1}{2} \log m$.
The area, or rather, perimeter law of the FQH states
has been the target of several recent studies
\cite{schoutens,friedman,fradkin}, in order
to confirm its validity and to compute
the value of $\gamma$ predicted in \cite{kitaev}
(see \cite{latorre2,schoutens} for the study of the entanglement
entropy for particle partitioning).
Reference \cite{fradkin} uses Chern-Simons theory, finding
the predicted value of
$\gamma$, however the linear behaviour of $S_A$,
is not captured,  due to the
purely topological nature of this theory. There are  numerical
studies using the Laughlin wave function \cite{schoutens} and exact
diagonalization \cite{friedman}, for filling fractions
$\nu = 1/3, 1/5$ and the $\nu = 5/2$  Pfaffian state.
The approaches of \cite{schoutens,friedman}
use the orbital basis for the Landau levels.
The close relationship of this basis to the spatial partitioning
of the blocks leads to an area law of the form
$S_A = c \sqrt{l_A} - \gamma + \CO(1/l_A)$, where
$l_A$ is the number of Landau orbitals in the block $A$.
The numerical values of $\gamma$ computed in the spherical
geometry \cite{schoutens} and the torus geometry \cite{friedman}  agree,
within some precision, with their theoretical values, despite
of the fact that the systems analyzed are not very large. We remark that the previous form of the area law in the orbital basis holds only in the case of fractional fillings. For integer fillings the orbital partitioning entropy is actually zero since the ground state is simply a product state in that basis.

In this letter we address the problem of computing the
entanglement entropy $S_A$ directly in real space, for
the Integer Quantum Hall states with $\nu \geq 1$,
in three different domains: strips in the torus, annulus in
the disk and casquettes in the sphere.
The reason for choosing integer filling fractions
is that the ground state is given by free fermions,
where standard techniques for computing entanglement entropies
are available \cite{latorre}. We find the area
law  $S_A \approx c_\nu L - \gamma$, with $\gamma =0$
in agreement with general arguments \cite{kitaev}. The non universal
constant $c_\nu$ is computed analytically for $\nu =1$
and numerically up to $\nu = 5$.
We also analyze the crossover from thin to large blocks,
finding that the onset of the area law occurs
when the width of the boundary is larger that a correlation length.
The blocks,  whose entropy we have computed,  are adapted
to the standard gauge choices used to analyze
the geometries of the torus, the disk and the sphere.

Let us consider the Landau model for a particle in a torus
of size $L_x \times L_y$. The one particle wave function in the lowest
Landau level (LLL),  in the gauge ${\bf A} = B(0,x)$,  is
(in units of the magnetic length $\ell $ equal to one):
\beq
\phi_{k_y}(x,y) = \frac{1}{ \pi^{1/4} L_y^{1/2}}  \; e^{i k_y y}
\; e^{- (x- k_y)^2/2} .
\label{1}
\eeq
On the torus, the identification of the wave function along the
$y$ direction implies:
\beq
k_y = \frac{2 \pi n}{L_y}, \quad   - \frac{n_0}{2} + 1 \leq n \leq
\frac{n_0}{2} .
\label{2}
\eeq
The number of LLLs, $n_0$, is obtained imposing that the particle
lives in the strip $|x| \leq L_x/2$, which yields
$n_0 = \frac{L_x L_y}{2 \pi}$.
%
%
%
%
This value also gives the total number of quantum fluxes
through the box.
The electron operator can be written as
\beq
\psi(x,y) = \sum_{k_y} \phi_{k_y}(x,y)  c_{k_y} + {\rm higher} \; {\rm LLs} ,
\label{6}
\eeq
where $c_{k_y}$ is the fermionic destruction
operator of the LLL labelled by $k_y$.
The extra term in (\ref{6}) involves the remaining Landau
levels, which are empty for filling fraction $\nu = 1$. Later on,
we shall take them into account when considering
higher filling fractions $\nu$.
The ground state for $\nu = 1$ is given by:
\beq
|\Phi_0 \rangle = \Pi_{k_y} c_{k_y}^\dagger  |0 \rangle ,
\label{7}
\eeq
where $|0 \rangle $ is the Fock vacuum.
The two point fermion correlator  in this state is,
\beq
C_{\br, \br'} = \langle \Phi_0 | \psi^\dagger(x,y)
\; \psi(x',y') | \Phi_0  \rangle .
\label{8}
\eeq
Using (\ref{6}) and (\ref{7}) one finds:
\beq
C_{\br, \br' } = \sum_{k_y}  \phi_{k_y}^*(x,y) \; \phi_{k_y}(x',y') ,
\label{9}
\eeq
and plugging (\ref{1})
\beq
C_{\br, \br' } =   \frac{1}{ \pi^{1/2} L_y}  \sum_{k_y}
  \; e^{i k_y ( y' - y) } \; e^{-  \frac{1}{2} ( (x- k_y)^2 + (x' - k_y)^2)} .
\label{10}
\eeq
The sum in (\ref{10})
is over the $n_0$ values of $k_y$ given in (\ref{2}).
In the limit $L_x, L_y \rightarrow \infty$ the correlator
(\ref{9}) becomes
\beq
C_{\br, \br' } =   \frac{1}{ 2 \pi }
e^{-  \frac{1}{4} (x- x' )^2  -  \frac{1}{4} (y- y' )^2
-  \frac{i}{2} (x+x' ) (y - y')    } ,
\label{11}
\eeq
and it is short range with a correlation
length proportional to the magnetic length $\ell = 1$.

We want to  compute the entanglement entropy, $S_\CD$,  of the
state $\Phi_0$,  in the strip
\beq
\CD : - \frac{l_x}{2} \leq x \leq \frac{l_x}{2}, \quad 0 \leq y \leq L_y .
\label{12}
\eeq
This  entropy  is given by the formula $S_\CD = {\rm Tr}_{\CD_c}
|\Phi_0 \rangle \langle \Phi_0|$, where $\CD_c$ is the complement
of $\CD$ in the torus.  The computation of $S_\CD$
is done in two steps \cite{latorre}. First one restricts the
correlation matrix
$C_{\br, \br' }$, to the domain $\CD$, i.e.
\beq
\tilde{C}_{\br, \br' } =  C_{\br, \br' }, \quad \br, \br' \in \CD .
\label{13}
\eeq
Next, one diagonalizes $\tilde{C}_{\br, \br' }$, i.e.
\beq
\int_{\CD} d^2 \br' \; \tilde{C}_{\br, \br' } \; f_m(\br' )
= \lambda_m f_m(\br) .
\label{14}
\eeq
The  entropy $S_\CD$ is  obtained
by means of,
\beq
S_\CD = \sum_m  H(\lambda_m) ,
\label{15}
\eeq
where $H(x) = - x \log x - (1-x) \log(1-x)$.
The eigenvalue problem (\ref{14}) can be rather
difficult for a generic domain $\CD$, however
for the strip (\ref{12}) this taks simplifies.
The basic observation is that  $\tilde{C}_{\br, \br' }$
only depends on the difference $y - y'$, which
suggests the ansatz
\beq
f_m(\br) = e^{ - i \mu_m y} g_m(x) .
\label{16}
\eeq
Plugging (\ref{16})  into (\ref{14}), and taking the limit
$L_y \rightarrow \infty$ one gets
\beq
e^{-  \frac{1}{2} x^2  +  \mu_m  x - \mu^2   } A_m
\;  = \lambda_m \;  g_m(x) ,
\label{17}
\eeq
where
\beq
A_m = \int_{-l_x/2}^{l_x/2}
\frac{dx}{ \pi^{1/2}}   e^{-  \frac{1}{2} x^2  +  \mu_m  x  }
\;  g_m(x) .
\label{18}
\eeq
For a non vanishing
eigenvalue $\lambda_m$, eq.(\ref{17})
fixes the function $g_m(x)$,  up to an overall factor.
Plugging (\ref{17}) into
(\ref{18}), the constant $A_m$ drops, and one gets
the eigenvalue
\beq
\lambda_m   =  \int_{-l_x/2}^{l_x/2}   \frac{dx}{ \pi^{1/2}}
e^{-   (x - \mu_m)^2  } .
\label{19}
\eeq
On the other hand, if $\lambda_m =0$ ,
eq.(\ref{17}) yields $A_m =0$, which becomes
a condition for the function $g_m$. However,
vanishing eigenvalues do not contribute to the entropy
(\ref{15}), so the solution of  $A_m =0$ is not required.
Recalling that the function (\ref{16})
is defined on the domain (\ref{12}),  one obtains
a quantization condition similar to eq.(\ref{2})
\beq
\mu_m = \frac{2  \pi m}{L_y}, \quad,  - \frac{n_0}{2} + 1 \leq m \leq
\frac{n_0}{2} .
\label{20}
\eeq
In fact, the eigenfunctions $f_m$ of $\tilde{C}_{\br, \br'}$
coincide with the conjugate of the LLL eigenfunctions
$\phi_{k_y}^*$, under the identification
$k_y = \mu_m$. Moreover, eq. (\ref{19})
can be written as the norm of (\ref{1}) over the
domain (\ref{12}), i.e.
\beq
\lambda_m   =  \int_\CD    d^2 \br \;  |\phi_{\mu_m}( \br)|^2 ,
\label{19b}
\eeq
which means that $\lambda_m$ is the probability
of finding the electron in the state $k_y = \mu_m$
in the domain $\CD$.
Integrating (\ref{19}) yields
\beq
\lambda_m \equiv \lambda(\mu_m, l_x)   = \frac{1}{2}
\left[ \Erf( \mu_m + \frac{l_x}{2}) -
\Erf( \mu_m - \frac{l_x}{2}) \right] ,
\label{21}
\eeq
where $\Erf(x)$ is the error function.
The function $H(\lambda(\mu,l_x))$
is localized in the regions $|\mu| \sim l_x/2$, associated
to the boundaries of $\CD$, where it can be approximated as
%
%
%
\beq
\lambda(\mu, l_x)    \sim   \frac{1}{2}
\left[ 1 - \Erf( |\mu| - \frac{l_x}{2}) \right]
=  \frac{1}{2} \Erfc( |\mu| - \frac{l_x}{2}) ,
\label{22}
\eeq
where $\Erfc(x) = 1 - \Erf(x)$ is the complementary error function.
In the limit $L_y >>1$, one can use eq. (\ref{20}) to
write (\ref{15}) as the integral,
\beq
S_\CD\equiv S(l_x, L_y) =  \frac{L_y}{2 \pi}
\int_{ - \infty}^\infty d \mu \; H(\lambda(\mu, l_x)) .
\label{23}
\eeq
Furthermore,  if $l_x >>1$, the main contribution to
(\ref{23}) comes from  the values of $\mu$ around $ \pm l_x/2$,
where one can use the approximation (\ref{22}). Shifting the integration
variable $\mu$,  one finally obtains
\beq
S(l_x, L_y) =  2 c_{\rm torus} \;    L_y ,
\label{24}
\eeq
where the constant $c_{\rm torus}$ is given by
\beq
c_{\rm torus} =
 \int_{ - \infty}^\infty  \frac{d \mu}{2 \pi}  \;
H(  \frac{1}{2} \Erfc( \mu)  ) \approx   0.20329081 .
\label{25}
\eeq
Equation (\ref{24}) is the area law satisfied by
$S_\CD$ in the limit  $l_x >>1$.
The absence of a constant
term in (\ref{24}), implies the vanishing of
the topological entropy $\gamma$  \cite{kitaev}.

\begin{figure}[t!]
\begin{center}
\hbox{\includegraphics[height= 4.1cm]{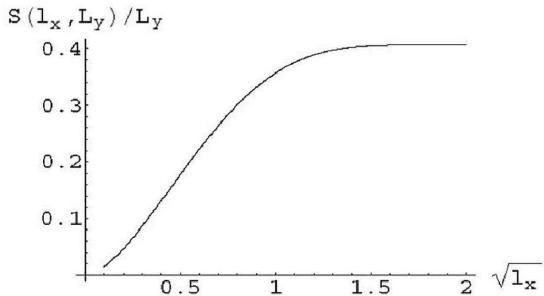}}
\end{center}
\caption{Plot of $S(l_x,L_y)$ as a function of $\sqrt{l_x}$ for
$L_y=20, 30, 40$.}
\label{entro-lx}
\end{figure}

Equations (\ref{24}) and (\ref{25})  were
derived under the asumption that $l_x >>1$,
but using eq.(\ref{15}) one can  estimate the value of $l_x$
above which the area law (\ref{24}) starts to be valid.
In fig. \ref{entro-lx},  we plot $S(l_x, L_y)/L_y$
as a function of $\sqrt{l_x}$, for various values of $L_y$.
For  $l_x \geq l_{x,c} = 2$ the entropy reaches a constant
value $2 c_{\rm torus}$
given by equation (\ref{25}), which agrees with the area law
(\ref{24}).
The value of $l_{x,c}$ is easy
to understand, it means that the width of $\CD$, along the $x$ direction,
must be bigger than two magnetic lengths, which guarantees
the existence of two boundaries with a width of
one correlation length each. Fig. \ref{entro-lx}
also shows that  the entropy increases linearly with
$\sqrt{l_x}$ for $l_x < l_{x,c}$, and reminds the result
of reference \cite{friedman}, where the entropy
of the $\nu =1/3, 1/5$ FQH states was computed for the
torus in the orbital basis. These authors found
that the entropy of a strip of width $\ell$,
varies as $\sqrt{\ell}$, where $\ell$ label the orbital
angular momenta. Our computation and that of reference
\cite{friedman} differ both in the basis and the filling
fraction, so the previous comparison has to be taken with care.

Let us now consider the disk geometry in the Landau gauge
where the eigenfunctions of the LLL are given by
\beq
\phi_{m}(z) = \frac{ z^m}{ (2 \pi 2^m m!)^{1/2}}
\; e^{- |z|^2/2}, \quad m =0,1, \dots .
\label{26}
\eeq
\no
The two point fermion correlator is similar to eq.(\ref{11}),
\beq
C_{z, z' } =  \frac{1}{2 \pi} e^{ - \frac{1}{4} ( |z|^2 + |z'|^2 + 2 z^* z')} .
\label{27}
\eeq
We want to compute the entanglement entropy in the annulus of radii
$r_1 < r_2$
\beq
\CD:   r_1 < |z| < r_2 .
\label{28}
\eeq
The procedure follows closely the case of the torus.
The eigenfunctions $f_m(z)$
of eq.(\ref{14}), with  non zero eigenvalues $\lambda_m$,
are given by $f_m(z) = \phi_m^*(z) \; (m=0,1, \dots)$
and the eigenvalues are
\beq
\lambda_m =  \int_{\CD} d^2 z \; |\phi_m(z)|^2, \;\;
m=0, 1 , \dots .
\label{29}
\eeq
Plugging (\ref{26}) into (\ref{29})
and performing the integral over the domain
(\ref{28}),  one finds
\beq
\lambda_m(r_1, r_2)  = \frac{1}{m!} \left[ \Gamma(m+1,\frac{r_1^2}{2})
-  \Gamma(m+1,\frac{r_2^2}{2}) \right] .
\label{30}
\eeq
The entropy of the annulus  is given by
\beq
S(r_1, r_2) =  \sum_{m=0}^\infty H(\lambda_m(r_1, r_2)) ,
\label{31}
\eeq
which, for large values of $r_1$ and $r_2$, satisfies the
area law
\beq
S(r_1,r_2) =  2 \pi (r_1 + r_2) \; c_{\rm disk} ,
\label{32}
\eeq
with $c_{\rm disk} = c_{\rm torus}$.
Equation (\ref{32})
can be proved analytically along the same lines as was
done before.

Another example which can be solved explicitely
is that of an electron moving on a sphere
of radius $R$ under the influence of a radial
magnetic field created by a monopole at the origin.
In the gauge where the vector potential
is given by  ${\bf A}= \hbar Q/e R \cot \phi$,
the wave functions
of the LLLs are the monopole harmonics,
\beq
Y_{Q, Q,m} = \left( \frac{ 2 Q +1}{ 4 \pi}
\left(
\begin{array}{c}
2 Q \\
Q-m \\
\end{array}
\right) \right)^{1/2} \;
(-1)^{Q-m} u^{Q+m} \; v^{Q-m} ,
\label{33}
\eeq
where $u = \cos(\theta/2) e^{- i \phi/2}$,
 $v = \sin(\theta/2) e^{ i \phi/2}$, with
$\theta$ and $\phi$ the polar and azimuthal
angles,  and $2 Q$ the total quantum flux
traversing the sphere. The two point fermion correlator
is given by
\beq
C_{\br, \br' } = \sum_{m= -Q}^Q
Y_{Q, Q,m}^*(\theta, \phi) \; Y_{Q, Q,m}(\theta', \phi') .
\label{34}
\eeq
Making the change $n = m + Q$ one can write (\ref{34}) as
\beq
C_{\br, \br' } = \frac{2 Q +1}{4 \pi}
\sum_{n=0}^{2 Q} \left(
\begin{array}{c}
2 Q \\
n \\
\end{array}
\right)  ( \bar{u} u')^n \; ( \bar{v} v' )^{2 Q - n} .
\label{35}
\eeq
We are interested in computing the entanglement entropy
in the spherical segment (i.e. casquette)
\beq
\CD: \theta_a  < \theta  < \theta_b
\label{36}
\eeq
bounded by the polar angles $\theta_1$ and $\theta_2$.
The eigenfunctions
of the correlator (\ref{35}) in the domain (\ref{36}),
with non zero eigenvalue,
are given by  $f_m = Y^*_{Q,Q,m}$, with
\beq
\lambda_n = \int_{\theta_a}^{\theta_b}  d \theta \int_0^{2 \pi} d \phi
|Y_{Q,Q,n-Q}|^2 .
\label{37}
\eeq
Performing the integral one finds,
\barray
& \lambda_n(\theta_a, \theta_b)   =
 (2 Q + 1)   \left(
\begin{array}{c}
2 Q \\
n \\
\end{array}
\right) & \label{38} \\
& \times
\left[ B( \cos^2( \frac{ \pi a}{ 4 Q + 2} ), 1 + n, 1 - n + 2Q)
\right. &  \nonumber  \\
& -  \left.    B( \cos^2( \frac{ \pi b}{ 4 Q + 2} ), 1 + n, 1 - n + 2Q)
\right], \;\; (n=0,1,\dots, 2Q) , &
\nonumber
\earray
where $B(z,n,m)$ is the incomplete beta function
and $\theta_{a(b)} = a(b) \pi/(2 Q +1)$. The entropy
of the region (\ref{36}) is computed from
\beq
S(\theta_a, \theta_b) =  \sum_{m=0}^{2 Q}
H(\lambda_m(\theta_a,\theta_b)) .
\label{39}
\eeq
The perimeter of the casquette of angle $\theta$ is given
by $P_\theta = 2 \pi R \sin \theta$, where the radius is given by
$R = \sqrt{Q }$ as follows from computing the number of quantum
fluxes. For large values of $Q$,
the entropy (\ref{39}) satisfies the area law
\beq
S(\theta_a, \theta_b) = (P(\theta_a) + P(\theta_b)) \; c_{\rm sphere} ,
\label{40}
\eeq
with $ c_{\rm sphere} = c_{\rm torus} = c_{\rm disk}$,
so that the three geometries yield
the same entropy per unit length of the perimeter.

\begin{figure}[t!]
\begin{center}
\hbox{\includegraphics[height=5cm]{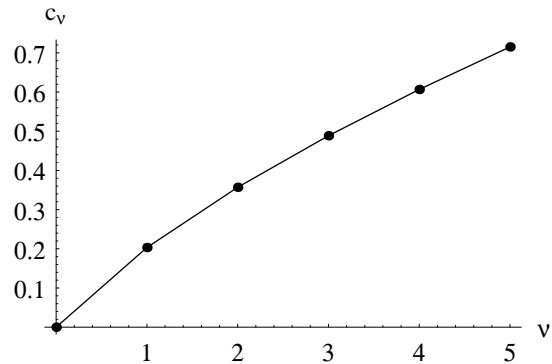}}
\end{center}
\caption{
The points denote the values of the constant
$c_\nu$ in the area law for integer
filling fractions $\nu=1, \dots, 5$. The continuous
line is a forth order polynomial fit.}
\label{c-versus-n}
\end{figure}

The previous results can be easily generalized
for integer filling fractions $\nu > 1$.
The correlation matrix $C_{\br, \br' }$ is given
by
\beq
C_{\br, \br' } = \sum_{n=0, \nu-1}
\sum_{m}  \phi_{n,m}^*(\br) \; \phi_{n,m}(\br') ,
\label{41}
\eeq
where  $\phi_{n,m}(\br)$ is the
wave function of the state $m$ in the
$n^{\rm th}$ Landau level.
The eigenfunctions of $\tilde{C}_{\br, \br' }$
are linear combinations of
$\phi_{n,m}^*(\br)$ with $n = 0, 1, \dots, \nu-1$,
and their eigenvalues are those of the
$\nu \times \nu$ matrix,
\beq
\Lambda_{m}(n,n')   =  \int_\CD    d^2 \br \;  \phi_{n,m}^*( \br)
 \phi_{n',m}( \br), \;\; n,n' =0, \dots, \nu-1 .
\label{42}
\eeq
The entanglement entropy is computed using eq.(\ref{15}),
where the summation runs over all the eigenvalues of
$\Lambda_m$. The area law $S_\CD \approx c_\nu L - \gamma_\nu$
remains valid, with $\gamma_\nu = 0$, and a value of $c_\nu$, which
depends on the filling fraction (see fig. \ref{c-versus-n}).

In summary, we have derived in this letter
the area law satisfied by the
entanglement entropy of the integer Quantum Hall states
with filling fraction, $\nu$,
for different types of domains in the torus,
the disk and the sphere. We have computed the
non universal constant $c_\nu$ of the area law as a function
of $\nu$. The topological entanglement
entropy vanishes, in agreement with the
theoretical results \cite{kitaev,fradkin}.
For $\nu = 1$, we have found a simple interpretation
of the area law. In this case the
entanglement entropy is given by the sum over the
LLL states, of the Shanon entropies associated
to finding an electron or a hole in the domain.
The area law arises from the contribution of
the LLLs inside a correlation length of the boundary
of the domain. Our method allows the computation
of the entanglement entropy for more complicated
domains. Of special interest are those with curvature
singularities wether one may expect deviations
from the area law.

{\it Acknowledgments-}
We thanks K. Schoutens and P. Calabrese for helpful conversations.
This work was supported by the  CICYT project FIS2004-04885 (G.S.).
I.R. also acknowledges an exchange grant of the ESF Science Programme
INSTANS 2005-2010. We thank the Galileo Galilei Institute for
Theoretical Physics for  the hospitality and the INFN
for partial support during the completion of this work.

\end{document}